\documentclass[12pt,preprint]{aastex}

\shorttitle{Heating of Cooling Flows}
\shortauthors{Fujita \& Suzuki}

\begin{document}

\title{On Heating of Cluster Cooling Flows by Sound Waves}

\author{Yutaka Fujita\altaffilmark{1,2} 
and Takeru Ken Suzuki\altaffilmark{3,4}}

\altaffiltext{1}{National Astronomical Observatory, Osawa 2-21-1,
Mitaka, Tokyo 181-8588, Japan; yfujita@th.nao.ac.jp.}

\altaffiltext{2}{Department of Astronomical Science, The Graduate
University for Advanced Studies, Osawa 2-21-1, Mitaka, Tokyo 181-8588,
Japan.}

\altaffiltext{3}{Department of Physics, Kyoto University, Kitashirakawa,
Sakyo-ku, Kyoto 606-8502, Japan.}

\altaffiltext{4}{JSPS Research Fellow.}

\begin{abstract}
 We investigate heating of the cool core of a galaxy cluster through the
 dissipation of sound waves excited by the activities of the central
 active galactic nucleus (AGN). Using a weak shock theory, we show that
 this heating mechanism alone cannot reproduce observed temperature and
 density profiles of a cluster, because the dissipation length of the
 waves is much smaller than the size of the core and thus the wave
 energy is not distributed to the whole core. However, we find that if
 it is combined with thermal conduction from the hot outer layer of the
 cluster, the wave heating can reproduce the observational results.
\end{abstract}


\section{Introduction}
\label{sec:intro}

The radiative cooling time of gas in the central regions of galaxy
clusters (cool cores) is generally much smaller than the Hubble time. In
the absence of any heating sources, this means that the intracluster
medium (ICM) flows subsonically toward the cluster center with a mass
deposition rate of $\dot{M}\sim 100$--$1000\: M_\sun\:\rm yr^{-1}$
\citep{fab94}. This flow was called a ``cooling flow''. However, recent
X-ray observations have shown that the cooling rate of the ICM is much
smaller. The Japanese {\it ASCA} team indicated that metal emission
lines from the low temperature cooling gas were much weaker than that had
been predicted by the classical cooling flow model
\citep{ike97,mak01}. This has been confirmed by {\it XMM-Newton}
\citep{pet01,kaa01,tam01}; the actual mass deposition rates are about
1/10 of those predicted by the classical cooling flow model.

The lack of metal emission lines indicates that the gas is prevented
from cooling by some heating sources. At present, the most popular
candidate for the heating source is the active galactic nucleus (AGN) at
the cluster center
\citep*{tuc83,rep87,bin95,cio01,boh02,chu02,rey02}. However, it is not
understood how the energy ejected by the AGN is transfered into the
surrounding ICM. One idea is that bubbles inflated by AGN jets move
outward in a cluster by buoyancy and mix the surrounding ICM
\citep*{chu01,qui01,sax01}. As a result of the mixing, hot ICM in the
outer region of the cluster is brought into, and subsequently heats, the
cluster center. The other idea is that the dissipation of sound waves
created through the AGN activities. In fact, sound waves or weak shocks
that may have evolved from sound waves are observed in the Perseus and
the Virgo clusters \citep{fab03a,for05}. \citet{fab03a} and
\citet{fab05} argued that the viscous dissipation of the sound waves is
responsible for the heating of a cool core. They estimated the
dissipation rate assuming that the waves are linear. However, when the
amplitude of sound waves is large, the waves rapidly evolve into
non-linear weak shocks \citep{ss72}, and their dissipation can be faster
than the viscous dissipation of linear waves \footnote{Of course
viscosity plays an important role for formation and dissipation of
shocks. For simplicity, however, we refer to the energy dissipation of
linear sound waves through viscosity as `viscous dissipation', and refer
to that through formation of non-linear weak shocks as `shock
dissipation'.}. Although \citet{fab03a} argued the presence of weak
shocks, their evolution from sound waves was not
considered. \citet*{rus04} performed numerical simulations of
dissipation of sound waves created by AGN activities. Their results
actually showed that the sound waves became weak shocks. However, their
simulations were finished before radiative cooling became
effective. Thus, the long-term balance between heating and cooling is
still unknown. In this letter, we consider the evolution of sound waves
to weak shocks, and analytically estimate the `time-averaged' energy
flux of the propagating waves as a function of distance, explicitly
taking account of the dissipation at weak shock fronts and its global
balance with radiative cooling. We assume the Hubble constant of
$H_0=70\:\rm km\: s^{-1}\: Mpc^{-1}$.

\section{Models}

We assume that sound waves are created by central AGN activities. The
waves propagate in the ICM outwards. These waves, having a relatively
large but finite amplitude, eventually form shocks to shape sawtooth
waves \citep{lan59,mih84}. If the velocity amplitude is larger than
$\sim 0.1$ sound velocity (the Mach number is $\gtrsim 1.1$), those
waves steepen and become weak shocks after propagating less than a few
wavelengths \citep[e.g.][]{suz02}. These shock waves directly heat the
surrounding ICM by dissipating their wave energy. We adopt a heating
model for the solar corona based on a weak shock theory
(\citealp{suz02,ss72}, see also \citealp*{fuj04a}). We assume that a
cluster is spherically symmetric and steady.

The equation of continuity is

\begin{equation}
\label{eq:cont}
 \dot{M}=-4\pi r^2\rho v\:,
\end{equation}
where $\dot{M}$ is the mass accretion rate, $r$ is the distance from the
cluster center, $\rho$ is the ICM density, and $v$ is the ICM
velocity. The equation of momentum conservation is
\begin{equation}
\label{eq:motion}
 v\frac{dv}{dr} = -\frac{GM(r)}{r^2}-\frac{1}{\rho}\frac{dp}{dr}
-\frac{1}{\rho c_s \{1+[(\gamma+1)/2]\alpha_w\}}
\frac{1}{r^2}\frac{d}{dr}(r^2 F_w)
\end{equation}
where $G$ is the gravitational constant, $M(r)$ is the mass within
radius $r$, $p$ is the ICM pressure, $c_s$ is the sound velocity,
$\gamma (=5/3)$ is the adiabatic constant, and $\alpha_w$ is the wave
velocity amplitude normalized by the ambient sound velocity
($\alpha_w=\delta v_w/c_s$). The wave energy flux, $F_w$, is given by
\begin{equation}
\label{eq:Fw}
 F_w = \frac{1}{3}\rho c_s^3 
\alpha_w^2 \left(1+\frac{\gamma+1}{2}\alpha_w\right) \:.
\end{equation}
The energy equation is
\begin{equation}
\label{eq:energy}
\rho v \frac{d}{dr}\left(\frac{1}{2}v^2
+\frac{\gamma}{\gamma-1}\frac{k_B T}{\mu m_H}\right)
+\rho v \frac{G M(r)}{r^2}
+\frac{1}{r^2}\frac{d}{d r}[r^2(F_w+F_c)]
+n_e^2\Lambda(T)=0 \:,
\end{equation}
where $k_B$ is the Boltzmann constant, $T$ is the ICM temperature, $\mu
(=0.6)$ is the mean molecular weight, $m_H$ is the hydrogen mass, $n_e$
is the electron number density, and $\Lambda$ is the cooling
function. The term $\nabla\cdot\mbox{\boldmath $F$}_w$ indicates the
heating by the dissipation of the waves. We adopt the classical form of
the conductive flux for ionized gas,
\begin{equation}
\label{eq:Fc}
 F_c = -f_c \kappa_0 T^{5/2}\frac{dT}{dr}
\end{equation}
with $\kappa_0 = 5\times 10^{-7}$ in cgs units. The factor $f_c$ is the
ratio of actual thermal conductivity to the classical Spitzer
conductivity. We adopt the cooling function for the metalicity of
$Z=0.5\: Z_\sun$, which is the typical value in the central region of a
cluster,
\begin{equation}
\label{eq:cool}
 n_e^2 \Lambda(T) = [C_1 (k_B T)^\alpha 
+ C_2 (k_B T)^\beta + C_3]n_i n_e\:,
\end{equation}
where $n_i$ is the ion number density, and the units for $k_B T$ are
keV. The constants in equation (\ref{eq:cool}) are $\alpha=-0.447$,
$\beta=-0.232$, $C_1=0.947$, $C_2=-1.71$, and $C_3=0.922$, and we can
approximate $n_i n_e=0.70(\rho/m_H)^2$. The units of $\Lambda$ are
$10^{-22}\:\rm ergs\: cm^3$. This approximation reproduces the cooling
function calculated by \citet{sut93} within $\sim 10$~\% for
$0.03\lesssim T \lesssim 25$~keV. The equation for the time-averaged
amplitude of the shock waves is given by
\begin{equation}
\label{eq:wave}
 \frac{d\alpha_w}{dr}=\frac{\alpha_w}{2}\left[
-\frac{1}{p}\frac{dp}{dr}
-\frac{2(\gamma+1)\alpha_w}{c_s \tau}-\frac{2}{r}
-\frac{1}{c_s}\frac{d c_s}{dr}
\right]\;,
\end{equation}
where $\tau$ is the period of waves, which we assume to be constant
\citep{ss72,suz02}. The second term of the right side of
equation~(\ref{eq:wave}) denotes dissipation at each shock front. We
assume that the mass distribution of a cluster can be represented by the
NFW profile \citep*{nav97}, although recent studies suggest a little
steeper profiles \citep[e.g.][]{fuk97}:

\begin{equation}
\label{eq:NFW}
 M(r) \propto \left[\ln \left(1+\frac{r}{r_s}\right)
-\frac{r}{r_s (1+r/r_s)}
\right]\:,
\end{equation}
where $r_s$ is the characteristic radius of the cluster. 

\section{Results}

For parameters of our model cluster, we adopt the observational data of
the Perseus cluster \citep*{ett02}. We assume that $r_s=280$~kpc,
$M(r_{1000})=3.39\times 10^{14}\: M_\sun$, and $r_{1000}=826$~kpc, where
the mean density within $r_\Delta$ is $\Delta$ times the critical
density of the Universe. Waves are injected at the inner boundary
$r=r_0$, which should be close to the size of bubbles observed at
cluster centers. We assume that $\lambda_0=r_0$, where $\lambda_0$ is
the initial wavelength. If the waves are injected in a form of sound
waves with amplitude $0.1\lesssim\alpha_w < 1$, waves travel about
$\lambda_0$ before they become shock waves \citep{suz02}. Therefore, for
$r_0\leq r \leq r_0+\lambda_0=2\lambda_0$, we assume that $\nabla \cdot
\mbox{\boldmath $F$}_w=0$ (eqs.~[\ref{eq:motion}]
and~[\ref{eq:energy}]), and that the second term of the right-hand side
of equation~(\ref{eq:wave}) is zero. The temperature, electron density,
and wave amplitude at $r=r_0$ are $T_0$ $n_{e0}$, and $\alpha_{w0}$,
respectively. Unless otherwise mentioned, the first two are fixed at
$T_0=3$~keV and $n_{e0}=0.08\:\rm cm^{-3}$, respectively, based on the
observational results of the Perseus cluster \citep{san04}.

\begin{figure}
\epsscale{1.0} \plotone{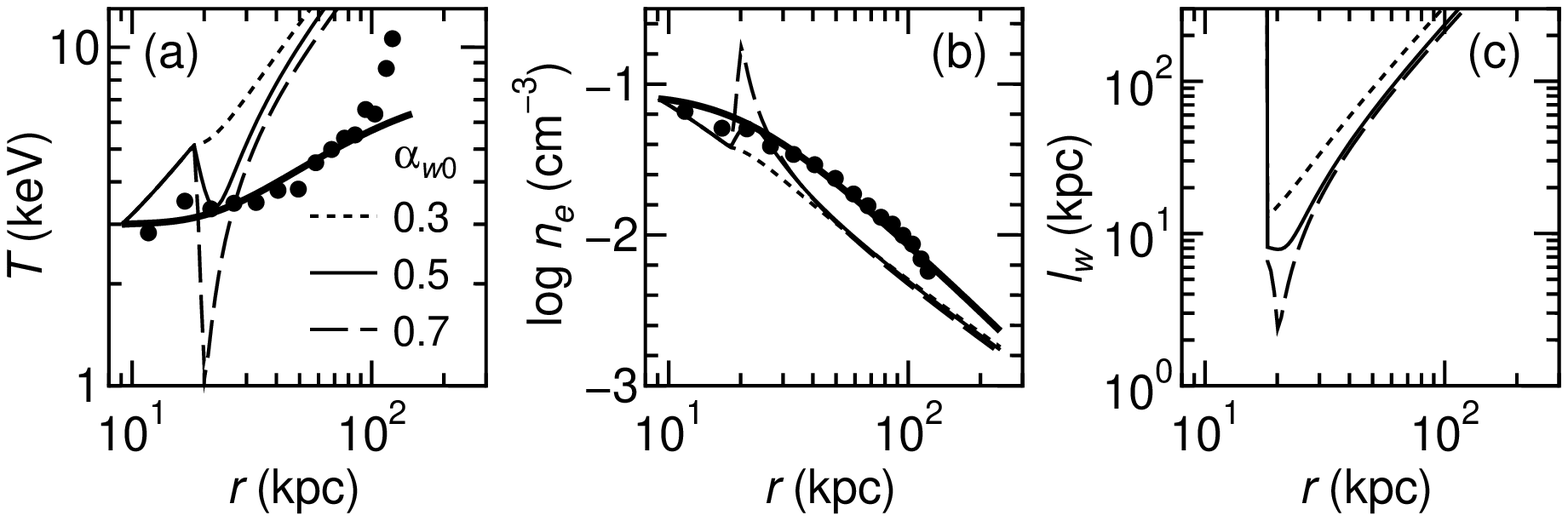} \caption{(a) Temperature, (b) density,
 and (c) dissipation length profiles for $\alpha_{w0}=0.3$ (dotted lines),
 0.5 (thin solid lines), and 0.7 (dashed lines). Other parameters are
 $\tau=1\times 10^7$~yr, $\dot{M}=50\: M_\sun\:\rm yr^{-1}$, and
 $f_c=0$. Filled circles are {\it Chandra} observations of the Perseus
 cluster \citep{san04}. The bold solid lines correspond to a genuine
 cooling flow model of $\dot{M}=500\: M_\sun\:\rm
 yr^{-1}$. \label{fig:alpha}}
\end{figure}

In Figure~\ref{fig:alpha}, we show the results when $\tau$ is fixed at
$1\times 10^7$~yr, and $\alpha_{w0}$ is changed. For the Perseus
cluster, \citet{fab03a} estimated that $\alpha_{w0}\sim 0.5$. The
dissipation length is defined as $l_w=|F_w/\nabla \cdot \mbox{\boldmath
$F$}_w|$. For these parameters, the initial wavelength is
$\lambda_0=9$~kpc, which is roughly consistent with the {\it Chandra}
observations \citep{fab03a}. The wave energy injection rate is given by
$\sim 4 \pi r_0^2 F_w(r_0)\sim 10^{45}\rm\: erg\: s^{-1}$, which is
comparable to the jet power of the nucleus in the Perseus cluster
\citep{fab02}. Other parameters are $\dot{M}=50\: M_\sun\:\rm yr^{-1}$,
and $f_c=0$. In general, larger $\dot{M}$ reproduces observed
temperature and density profiles better. However, large $\dot{M}$ is
inconsistent with recent X-ray observations as was mentioned in
\S~\ref{sec:intro}. For comparison, we show the results of a genuine
cooling flow model ($\dot{M}=500\: M_\sun\:\rm yr^{-1}$,
$\alpha_{w0}=0$, and $f_c=0$) and the {\it Chandra} observations of the
Perseus cluster \citep{san04}.  Figures~\ref{fig:alpha}a
and~\ref{fig:alpha}b show that only a small region is heated.  The jumps
of $T$ and $n_e$ at $r=2\lambda_0=18$~kpc are produced by weak shock
waves that start to dissipate there. The energy of the sound waves
rapidly dissipates at the shocks, which is clearly illustrated in short
dissipation lengths, $l_w\sim 2$--$15$~kpc
(Fig.~\ref{fig:alpha}c). These dissipation lengths are smaller than
those of viscous dissipation for linear waves, which can be represented
by $l_v=420 \:\lambda_{9}^2\: n_{0.08}\: T_3^{-2}$~kpc, where the
wavelength $\lambda = 9\:\lambda_9$~kpc, the density $n=0.08\:
n_{0.08}\rm\: cm^{-3}$, and the temperature $T=3\: T_3$~keV
\citep{fab03a,lan59}. In Figure~\ref{fig:alpha}, the ICM density becomes
large and the temperature becomes small at $r\gtrsim 2\lambda_0$ so that
the rapid shock dissipation is balanced with radiative cooling. Because
of this, waves cannot reproduce the observed temperature and density
profiles that gradually change on a scale of $\sim 100$~kpc. Note that
the density peaks in Figures~\ref{fig:alpha}b indicates that the
solutions are unstable. In an actual cluster, this would lead to
convection, and the ICM would be heated through the convection rather
than the dissipation of sound waves alone.

\begin{figure}
\epsscale{1.0} \plotone{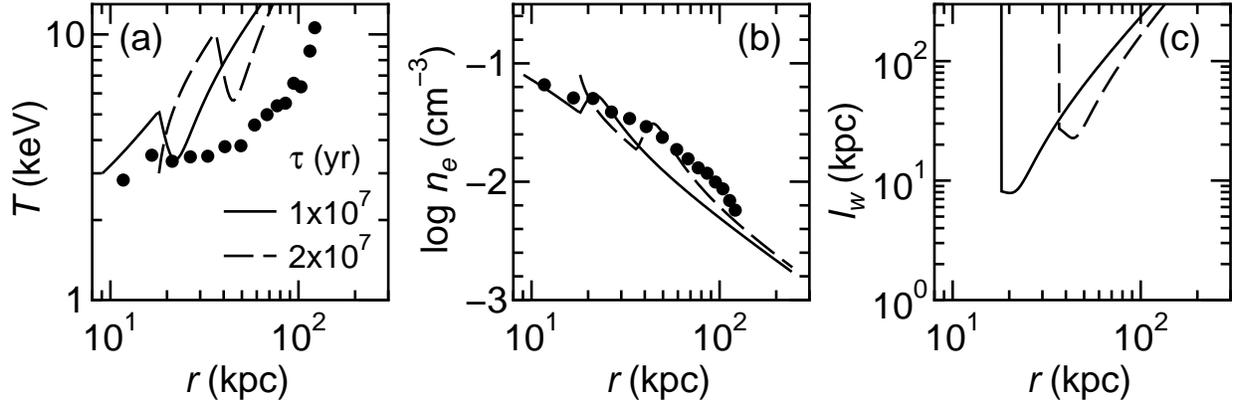} \caption{Same as Fig.~\ref{fig:alpha}
 but for $\tau=1\times 10^7$~yr (solid lines), and $2\times 10^7$~yr
 (dashed lines). Other parameters are $\alpha_{w0}=0.5$, $\dot{M}=50\:
 M_\sun\:\rm yr^{-1}$, and $f_c=0$. \label{fig:tau}}
\end{figure}

In Figure~\ref{fig:tau}, we present the results when $\tau=2\times
10^7$~yr. Compared with the case of $\tau=1\times 10^7$~yr, the wave
energy dissipates in outer regions. However, the dissipation lengths are
still smaller than the cluster core size ($\sim 100$~kpc). Note that
larger $\tau$ (or $\lambda_0$) means formation of larger bubbles. As
indicated by \citet{chu00}, it is unlikely that the size of the bubbles
becomes much larger than 20~kpc; the bubbles start rising through
buoyancy before they become larger. On the other hand, when
$\tau<10^7$~yr ($\lambda_0<9$~kpc), the waves heat only the ICM around
the cluster center. The predicted temperature and density profiles are
obviously inconsistent with the observations.

\begin{figure}
\epsscale{1.0} \plotone{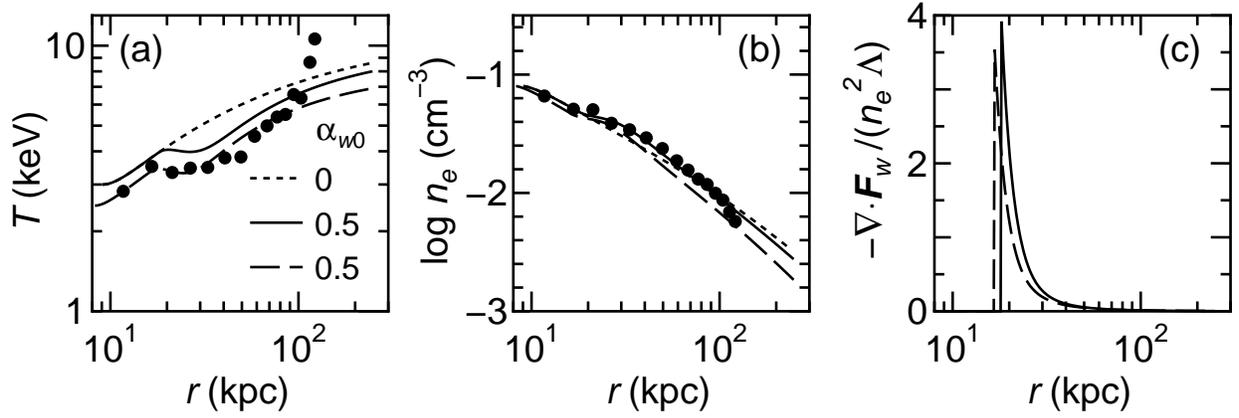} \caption{(a) Temperature, (b) density,
 and (c) dissipation strength profiles for $\alpha_{w0}=0$ and $T_0=3$~keV
 (dotted lines), $\alpha_{w0}=0.5$ and $T_0=3$~keV (solid lines), and
 $\alpha_{w0}=0.5$ and $T_0=2.5$~keV (dashed lines). Other parameters are
 $\tau=1\times 10^7$~yr, $\dot{M}=50\: M_\sun\:\rm yr^{-1}$, and
 $f_c=0.2$. Filled circles are {\it Chandra} observations of the Perseus
 cluster \citep{san04}. \label{fig:fc}}
\end{figure}

The inclusion of thermal conduction changes the situation
dramatically. Figure~\ref{fig:fc} shows the results when $f_c=0.2$. The
value of $f_c$ is based on the study of \citet{nar01}. The models
including both wave heating and thermal conduction can well reproduce
the observed temperature and density profiles. Figure~\ref{fig:fc}c
shows the contribution of the wave heating ($-\nabla \cdot
\mbox{\boldmath $F$}_w$) to compensating radiative cooling ($n_e^2
\Lambda$). Since $-\nabla \cdot \mbox{\boldmath $F$}_w/n_e^2 \Lambda >
1/2$ for $r\sim 20$--30~kpc, the wave heating is more effective than the
thermal conduction in that region.

\section{Discussion}

We showed that sound waves created by the central AGN alone cannot
reproduce the observed temperature and density profiles of a cluster,
because the dissipation length of the waves is much smaller than the
size of a cluster core and the waves cannot heat the whole core. The
same problem has been known for models of solar-corona heating by sound
waves \citep[e.g.][]{ss72}. The problem of the short dissipation length
should also be studied for other heating models including the bubble
motion in clusters. On the other hand, we found that if we include
thermal conduction from the hot outer layer of a cluster with the
conductivity of 20\% of the Spitzer value, the observed temperature and
density profiles can be reproduced. The idea of the ``double heating''
(AGN plus thermal conduction) was proposed by \citet{rus02}.

However, the fine structures observed in cluster cores may show that the
actual conductivity is much smaller than that we assumed
\citep[e.g.][]{fuj02, maz03}; the structures would soon be erased, if
the conductivity is that large. If the conductivity is small, we need to
consider other possibilities. While we considered successive minor AGN
activities, some authors consider that rare major AGN activities should
be responsible for heating of cool cores \citep{sok01,kai03}. In this
scenario, powerful bursts of the central AGN excite strong shocks and
heat the surrounding gas in the inner region of a cluster on a timescale
of $\gtrsim 10^{9}$~yr. In fact, \citet{mcn05} found such a violent
activity in a distant cluster. Moreover, in this scenario, heating and
cooling are not necessarily balanced at a given time, although they must
be balanced on a very long-term average. This is consistent with the
fact that there is no correlation between the masses of black holes in
the central AGNs and the X-ray luminosities of the central regions of
the clusters \citep{fuj04b}. Alternative idea is that cluster mergers
are responsible for heating of cool cores \citep*{fuj04,fuj05}. In this
``tsunami'' model, bulk gas motions excited by cluster mergers produce
turbulence in and around a core, because the cooling heavy core cannot
be moved by the bulk gas motions, and the resultant relative gas motion
between the core and the surrounding gas induces hydrodynamic
instabilities. The core is heated by turbulent diffusion from the hot
outer region of the cluster. Since the turbulence is produced and the
heating is effective only when the core is cooling and dense,
fine-tuning of balance between cooling and heating is alleviated for
this model.

\acknowledgments

Y.~F.\ was supported in part by a Grant-in-Aid from the Ministry of
Education, Culture, Sports, Science, and Technology of Japan
(17740162). T.~K.~S. is financially supported by the JSPS Research
Fellowship for Young Scientists, grant 4607.

\clearpage

\clearpage

\clearpage

\clearpage

\begin{thebibliography}{}

\bibitem[Binney \& Tabor(1995)]{bin95} Binney, J.,~\& Tabor, 
G.\ 1995, \mnras, 276, 663

\bibitem[B{\" o}hringer et al.(2002)]{boh02} B{\" o}hringer, 
H., Matsushita, K., Churazov, E., Ikebe, Y., \& Chen, Y.\ 2002, 
\aap, 382, 804 

\bibitem[Churazov et al.(2001)]{chu01} Churazov, E., Br{\" 
u}ggen, M., Kaiser, C.~R., B{\" o}hringer, H., \& Forman, W.\ 2001, \apj, 
554, 261 

\bibitem[Churazov et al.(2000)]{chu00} Churazov, E., Forman, 
W., Jones, C., \& B\"{o}hringer, H.\ 2000, \aap, 356, 788 

\bibitem[Churazov et al.(2002)]{chu02} Churazov, E.,
			       Sunyaev, R., Forman,
W., \& B{\" o}hringer, H.\ 2002, \mnras, 332, 729 

\bibitem[Ciotti \& Ostriker(2001)]{cio01} Ciotti, L.,~\& 
Ostriker, J.~P.\ 2001, \apj, 551, 131 

\bibitem[Ettori et al.(2002)Ettori, De Grandi, 
\& Molendi]{ett02} Ettori, S., De Grandi,
S., \& Molendi, S.\ 2002, \aap, 391, 841 

\bibitem[Fabian(1994)]{fab94} Fabian, A.~C.\ 1994, \araa, 32, 
277 

\bibitem[Fabian et al.(2002)]{fab02} Fabian, A.~C., Celotti, 
A., Blundell, K.~M., Kassim, N.~E., \& Perley, R.~A.\ 2002, \mnras, 331, 
369 
 
\bibitem[Fabian et al.(2005)]{fab05} Fabian, A.~C., Reynolds, C.~S.,
			       Taylor, G.~B., \& Dunn, R.~J.~H.\  2005,
			       submitted to \mnras\ (astro-ph/0501222)

\bibitem[Fabian et al.(2003)]{fab03a} Fabian, A.~C., Sanders, 
J.~S., Allen, S.~W., Crawford, C.~S., Iwasawa, K., Johnstone, R.~M., 
Schmidt, R.~W., \& Taylor, G.~B.\ 2003, \mnras, 344, L43 

\bibitem[Forman et al.(2005)]{for05} Forman, W. et al.\ 2005, \apj, in
			       press (astro-ph/0312576)

\bibitem[Fujita et al.(2004)Fujita, Matsumoto, \& Wada]{fuj04} Fujita,
Y., Matsumoto, T., \& Wada, K.\ 2004, \apjl, 612, L9

\bibitem[Fujita et al.(2005)]{fuj05} Fujita, Y., Matsumoto, 
T., Wada, K., \& Furusho, T.\ 2005, \apjl, 619, L139 

\bibitem[Fujita \& Reiprich(2004)]{fuj04b} Fujita, Y., \& 
Reiprich, T.~H.\ 2004, \apj, 612, 797 

\bibitem[Fujita et al.(2002)]{fuj02} Fujita, Y., Sarazin, 
C.~L., Kempner, J.~C., Rudnick, L., Slee, O.~B., Roy, A.~L., Andernach,
			       H.,
\& Ehle, M.\ 2002, \apj, 575, 764 

\bibitem[Fujita et al.(2004)Fujita, Suzuki, \& Wada]{fuj04a} Fujita, Y.,
			       Suzuki,
T.~K., \& Wada, K.\ 2004, \apj, 600, 650 

\bibitem[Fukushige \& Makino(1997)]{fuk97} Fukushige, T., \& 
Makino, J.\ 1997, \apjl, 477, L9 

\bibitem[Ikebe et al.(1997)]{ike97} Ikebe, Y., et al.\ 1997, 
\apj, 481, 660 

\bibitem[Kaastra et al.(2001)]{kaa01} Kaastra, J.~S., 
Ferrigno, C., Tamura, T., Paerels, F.~B.~S., Peterson, J.~R., \& Mittaz, 
J.~P.~D.\ 2001, \aap, 365, L99 

\bibitem[Kaiser \& Binney(2003)]{kai03} Kaiser, C.~R., \& 
Binney, J.\ 2003, \mnras, 338, 837 

\bibitem[Landau \& Lifshitz(1959)]{lan59} Landau, L. D., \& Lifshitz,
		    E. M.\ 1959, Fluid Mechanics, \S~94, \S~114 (London:
Pergamon)

\bibitem[Makishima et al.(2001)]{mak01} Makishima, K., et 
al.\ 2001, \pasj, 53, 401 

\bibitem[Mazzotta et al.(2003)]{maz03} Mazzotta, P., Edge, 
A.~C., \& Markevitch, M.\ 2003, \apj, 596, 190 

\bibitem[McNamara et al.(2005)]{mcn05} McNamara, B.~R., 
Nulsen, P.~E.~J., Wise, M.~W., Rafferty, D.~A., Carilli, C., Sarazin, 
C.~L., \& Blanton, E.~L.\ 2005, \nat, 433, 45 

\bibitem[Mihalas \& Mihalas(1984)]{mih84} Mihalas, D., \&
Mihalas, W.~B.\ 1984, Foundations of Radiation Hydrodynamics, \S~5.3
			       (New
York: Oxford Univ. Press)

\bibitem[Narayan \& Medvedev(2001)]{nar01} Narayan, R., \& 
Medvedev, M.~V.\ 2001, \apjl, 562, L129 

\bibitem[Navarro et al.(1997)Navarro, Frenk, \& White]{nav97} Navarro,
J.~F., Frenk, C.~S., \& White, S.~D.~M.\ 1997, \apj, 490, 493

\bibitem[Peterson et al.(2001)]{pet01} Peterson, J.~R.~et 
al.\ 2001, \aap, 365, L104 

\bibitem[Quilis et al.(2001)Quilis, Bower, \& Balogh]{qui01} Quilis, V., 
Bower, R.~G., \& Balogh, M.~L.\ 2001, \mnras, 328, 1091 

\bibitem[Rephaeli(1987)]{rep87} Rephaeli, Y.\ 1987, \mnras, 
225, 851 

\bibitem[Reynolds et al.(2002)Reynolds, Heinz, \&
			       Begelman]{rey02} Reynolds,
C.~S., Heinz, S., \& Begelman, M.~C.\ 2002, \mnras, 332, 271 

\bibitem[Ruszkowski \& Begelman(2002)]{rus02} Ruszkowski, M., 
\& Begelman, M.~C.\ 2002, \apj, 581, 223 

\bibitem[Ruszkowski et al.(2004)Ruszkowski, 
Br{\" u}ggen, \& Begelman]{rus04} Ruszkowski, M.,
Br{\" u}ggen, M., \& Begelman, M.~C.\ 2004, \apj, 615, 675 

\bibitem[Sanders et al.(2004)]{san04} Sanders, J.~S., Fabian, 
A.~C., Allen, S.~W., \& Schmidt, R.~W.\ 2004, \mnras, 349, 952 

\bibitem[Saxton et al.(2001)Saxton, Sutherland, \& Bicknell]{sax01} 
Saxton, C.~J., Sutherland, R.~S., \& Bicknell, G.~V.\ 2001, \apj, 563,

\bibitem[Soker et al.(2001)]{sok01} Soker, N., White, R.~E., 
David, L.~P., \& McNamara, B.~R.\ 2001, \apj, 549, 832 

\bibitem[Stein \& Schwartz(1972)]{ss72}
Stein, R. F. \& Schwartz, R. A. 1972, \apj, 177, 807

\bibitem[Sutherland \& Dopita(1993)]{sut93} Sutherland, 
R.~S.~\& Dopita, M.~A.\ 1993, \apjs, 88, 253 

\bibitem[Suzuki(2002)]{suz02} Suzuki, T.~K.\ 2002, \apj,
578, 598

\bibitem[Tamura et al.(2001)]{tam01} Tamura, T.~et al.\ 2001, 
\aap, 365, L87 

\bibitem[Tucker \& Rosner(1983)]{tuc83} Tucker, W.~H.~\& 
Rosner, R.\ 1983, \apj, 267, 547 

\end{thebibliography}
\end{document}